\documentclass[11pt,fleqn]{article}   
\usepackage{latexsym}
\usepackage{amsmath}
\catcode`\@=11
\@addtoreset{equation}{section}
\def\theequation{\arabic{section}.\arabic{equation}}
\def\appendix{\renewcommand{\thesection}{\Alph{section}}\setcounter{section}{0}
              \renewcommand{\theequation}
            {\mbox{\Alph{section}.\arabic{equation}}}\setcounter{equation}{0}}

\def\maketitle{\thispagestyle{empty}\setcounter{page}0\newpage
                \renewcommand{\thefootnote}{\arabic{footnote}}
                  \setcounter{footnote}0}
\renewcommand{\thanks}[1]{\renewcommand{\thefootnote}{\fnsymbol{footnote}}
               \footnote{#1}\renewcommand{\thefootnote}{\arabic{footnote}}}

\renewcommand{\title}[1]{\begin{center}\Large\bf #1\end{center}\rm\par\bigskip}
\renewcommand{\author}[1]{\begin{center}\Large #1\end{center}}
\newcommand{\address}[1]{\begin{center}\large #1\end{center}}

\def\dinfn{\smallskip Dipartimento di Fisica, Universit\`a di Trento\\ 
                           and Istituto Nazionale di Fisica Nucleare,\\
                                   Gruppo Collegato di Trento, Italia}

\def\Idinfn{\address{\dinfn}}

\newcommand{\email}[1]{e-mail: \sl #1@science.unitn.it\rm}

\def\babs{\hrule\par\begin{description}\item{Abstract: }\it} 
\def\eabs{\par\end{description}\hrule\par\medskip\rm}
\renewcommand{\date}[1]{\par\bigskip\par\sl\hfill #1\par\medskip\par\rm}
 
\newcommand{\s}[1]{\section{#1}}

\newcommand{\ca}[1]{{\cal #1}}         
\def\beq{\begin{eqnarray}}    
\def\eeq{\end{eqnarray}}      
\def\ben{\begin{itemize}}    
\def\een{\end{itemize}}      
\def\at{\left(}               
\def\aq{\left[}               
\def\ag{\left\{}              
\def\ct{\right)}              
\def\cq{\right]}              
\def\cg{\right\}}             
\def\R{{\hbox{{\rm I}\kern-.2em\hbox{\rm R}}}}   
\def\H{{\hbox{{\rm I}\kern-.2em\hbox{\rm H}}}}   
\def\N{{\hbox{{\rm I}\kern-.2em\hbox{\rm N}}}}   
\def\C{{\ \hbox{{\rm I}\kern-.6em\hbox{\bf C}}}} 
\def\Z{{\hbox{{\rm Z}\kern-.4em\hbox{\rm Z}}}}   
\def\pa{\partial}
\def\al{\alpha}

\def\ga{\gamma}
\def\de{\delta}

\def\ka{\kappa}
\def\la{\lambda}

\def\si{\sigma}

\def\Si{\Sigma}
\def\Om{\Omega}

\def\g{{g}}
\def\gg{{\tilde{g}}}
 
\begin{document}

\hfill{\sl preprint - UTF 450}

\title{Black hole entropy from classical Liouville theory} 
\author{A. Giacomini\thanks{\email{giacomin}}, N. Pinamonti\thanks{\email{pinamont}}}
\Idinfn
\date{January 2003}
\babs
In this article we compute the black hole entropy by finding a classical central charge of the Virasoro
algebra of a Liouville theory using the Cardy formula. 
This is done by performing a dimensional reduction of the Einstein Hilbert action
with the ansatz of spherical symmetry and writing the metric in conformally flat form.
We obtain two coupled field equations. 
Using the near-horizon approximation the field equation for the conformal factor decouples. 
The one concerning the conformal factor is a Liouville equation, it posses the symmetry induced by a
Virasoro algebra. We argue that it describes the microstates of the black hole, namely the generators of this 
symmetry do not change the thermodynamical properties of the black hole.
\eabs
\s{Introduction}

One of the most important results in black hole physics is that black holes have an entropy given by the
famous  Bekenstein-Hawking  \cite{bek72, bek73, haw75} formula 
\beq
S=\frac{A_h}{4G},
\eeq
that relates the entropy of a black hole with the spatial area of its horizon $A_h$.
As it is a thermodynamics relation, it would be interesting to find a microscopic
interpretation of it.
The  Bekenstein-Hawking formula is obtained by using a semiclassical approach only, no details of a possible
quantum gravity were considered. So it would be interesting to count the degrees of freedom without using
details of any quantum theory of gravity.  
A very promising approach consists finding a central extension of an algebra of diffeomorphisms, like in
the work of Strominger \cite{strom98} used for the computation of the BTZ \cite{btz92} black hole entropy. 
This approach uses the result of Brown and Henneaux \cite{BrownHenn} on
the central extension of the diffeomorphism algebra that preserves the $AdS$ structure
at infinity. That is a particular case of the correspondence between $AdS_{d+1}$ spaces and conformal field theories living on the boundary \cite{mald98}.
But the use of local symmetries at infinity does not permit to distinguish a black hole from a star, apart from topology.
Since the central role is played by the horizon of the black hole, it seems more intuitive to use near-horizon
symmetries. This has been done by Carlip et al.   
\cite{carlip, solo99, lin99, park02}, and in other papers as well \cite{bbo99}.
This second approach seems to involve some technical difficulties especially for the Schwarzschild black hole 
\cite{park99, park00, solov00}. 
On the other hand, this way leads to difficulties in understanding  the physical
interpretation of the counted degrees of freedom.
We shall analyze this aspect of the theory performing a dimensional reduction, as in the work of Solodukhin
\cite{solo99} or in \cite{cpp02}, but with a little bit different approach. 
In two dimensions\footnote{See
also the work \cite{VanzCaldCate} for some aspect of 2D black hole and
\cite{CadMin,caca01} for asymptotic 2D
symmetries.} the metric tensor is conformally flat, 
therefore we gain a conformally flat metric by the dimensional reduction of 
the four-dimensional action with spherical symmetry and by a conformal transformation.
The curvature is embodied in the field that generates that transformation.
At the end we obtain a theory of two scalar fields \cite{cghs, alwi93, medv01, rus92}: the dilaton (i.e. the radius of
the  event horizon) and the Liouville field (the conformal factor of the metric) which propagate in flat
space. 
Notice that for the thermodynamical properties of black holes the relevant aspect
is the geometry of the $(r,t)$-plane, e.g. the temperature of the black hole is proportional to the surface
gravity computed in the $(r,t)$-plane. 
This can also be seen in the Euclidean approach to black hole thermodynamics \cite{gib76}. 
Now in our model  the geometry of the $(r,t)$-plane is encoded in the Liouville field. Our proposal is that
this field, and its fluctuations, are responsible for the black hole entropy, and the dilaton has to be thought as
fixed. On the other hand it is the geometry of the $(r,t)$-plane that is modified by the presence of a black hole,
namely the singularity leads to a divergent conformal factor.
Like in the work of Jackiw \cite{jac84}, we put our attention to the equations of motion of the conformal
factor,  and not on the action and  the dilaton.
At this stage we may consider a fixed dilaton, and consequently we may expand the dilaton potential near the horizon. 
We notice that the equations of motion that survive are the equation for the Liouville field, 
therefore we consider the action of a classical Liouville theory as the action of the black hole.\\ 
The Liouville theory has a computable classical central charge. 
In our dimensional-reduced model the central charge is proportional to the area of the horizon. 
Then the generator $L_0$ can be computed using a near-horizon approximation. 
Using then the Cardy \cite{cardy} formula we obtain the usual  Bekenstein-Hawking
entropy. 
We use a more dynamical approach, in the sense that we consider a solution of the equation of
motion, and some fluctuations around it. 
The reason that makes the relevant field  the  Liouville field has to be sought for in a possible quantum
theory of gravity, but further investigations on the subject are needed. 
The Virasoro algebra we shall find, generating conformal diffeomorphisms, preserves the horizon and the
surface gravity of the black hole with a bifurcate Killing horizon. If the black hole is extremal our
construction does not hold.

The work presented here is organized in the following way:
in the next section we consider the dimensionally reduced theory, and  we show that this leads to a
dilatonic two-dimensional theory. In the third section we present the near-horizon approximation and its
effect on the classical equations of motion.  We consider only fluctuations of the conformal
factor of the metric to describe the microstates of the black hole.
The form of the approximated equation of motion is found to be Liouville-like. 
In the fourth section for completeness we present the Virasoro algebra of the charges of the theory
and then, in the fifth
one, we compute explicitly the central charge and the generators in the case of a black hole with a bifurcate
Killing horizon fixing the energy of the model equal to the energy of the black hole (namely the ADM mass).
Through this way we recover the correct entropy. 

\s{Dimensional Reduction}

In this section we want to study gravity near the  horizon in the presence of spherical symmetry. 
This is widely known in the literature: We recall here the work concerning the black holes 
\cite{cghs, ns98a, ns98back}. 
The standard procedure consists of using symmetries to reduce the dimensions of the space.
Then we study the gravitational action, from a Lagrangian point of view, in the presence of spherical
symmetry. In particular we consider a black hole solution with a bifurcate Killing horizon. 
In this framework we consider only the radial degrees of freedom of the metric.
As is well known, dimensional reduction leads to Liouville-like theories, or in general to dilatonic
two-dimensional theories.
We shall show that the equations of motion are given in terms of two fields, the dilaton $\eta$ and 
$\rho$ called Liouville field. 
Let us start with the ansatz of spherical symmetry \cite{medv01}
\beq\label{2dmetricfree}
ds^2=\g_{ab}^{(2)} dx^a dx^b +\Phi^2 d\Om^2,
\eeq
where $\g_{ab}^{(2)}$ is the metric of a two-dimensional spacetime, $d\Om^2$ is the metric of a 
two-dimensional sphere with radius equal to one. 
The field  $\Phi$, that depends on $x_a$, represents the radius of this sphere. 
We consider now the action of the four dimensional metric $\g^{(4)}$,
\beq
I=\frac{1}{16\pi}\int \sqrt{-\ga ^{(4)}} \ca{R}[\g^{(4)}]
\eeq
with $G=1$. If we integrate out the angular degrees of freedom, we obtain
\beq\label{2ddilatonaction}
I=\frac{1}{4}\int \sqrt{-\g^{(2)}} \at 2(\nabla \Phi )^2 +\Phi ^2 \ca{R}[\g^{(2)}] +2\ct.
\eeq 
This is the action of an effective two-dimensional dilatonic theory of four dimensional gravity. 
We notice, at this level, that the form of the two-dimensional metric is arbitrary because we have
 used the spherical symmetry only.
This action was already studied in the literature, for example \cite{cghs,rus92,gid93,solo99,carta}.
In order to put the action in a more useful form, we redefine
\beq\label{connmetric}
\Phi ^2 = \eta,  \qquad \g^{(2)}_{ab}=\frac{1}{\sqrt{\eta}}\; \gg_{ab},
\eeq
obtaining the action of a dilatonic two-dimensional theory in the usual form
\beq\label{2dVaction}
I=\frac{1}{2}\int \sqrt{-\gg} \aq \frac{\eta}{2}\ca{R}[\gg] +V(\eta) \cq,
\eeq
with $V(\eta)=1/\sqrt{\eta}$. 
This equation (with general potential) describes dimensional reduced four-dimensional space, 
but also spaces with general dimension\footnote{We recall that the Jackiw
Teitelboim model \cite{jac84, tei84} (a three dimensional space) has a linear dilaton potential
$V(\eta)=\eta$,  the CGHS model \cite{cghs} has a constant potential $V(\eta)=\la$.}.
Before going on with the main subject of this work we shall consider briefly the classical theory.
The derivation of the equations of motion from the action (\ref{2dVaction}) is a standard procedure, 
they read:
\beq
\ca{R}[\gg]+2\pa_\eta V(\eta)=0, \qquad \nabla_a\pa_b\eta-\gg_{ab}\Box_\gg \eta+\gg_{ab}V(\eta)=0,
\eeq
where the former equation comes from the variation of the action with respect to the field $\eta$ and the
latter from 
the variation with respect to the two-dimensional metric $\gg$. 
These equations are given in terms of 
the dilatonic field $\eta$ and the two-dimensional metric $\gg$.
The dilaton $\eta$ is related to the radius of the two-dimensional sphere by means of the 
former equation in (\ref{connmetric}) while $\gg$ encodes the geometry of the $(r,t)$-plane. 
The presence of a black hole modifies the geometry of the $(r,t)$-plane in a strong way.
We argue that the relevant degrees of freedom for the thermodynamics behavior of
the black hole are related with the geometry of the $(r,t)$-plane.  

It is well known that in two dimensions the metric can always be written in the conformally flat form  
\beq\label{bame}
\gg_{ab}=e^{-2\rho}\gamma_{ab}
\eeq
where $\gamma=(-1,1)$ is the Minkowskian two-dimensional metric. Therefore the field $\rho$ describes
the geometry of the $(r,t)$-plane completely. 
The action (\ref{2dVaction}) becomes
\beq\label{Ieff}
I=\frac{1}{2} \int d^2x\at -\pa_a\eta\pa^a\rho+V(\eta)e^{-2\rho}\ct.
\eeq
As usual, the action written above requires the constraints $\de I/\de g^{ab}=\al T_{ab}=0$
to describe the same dynamics as the action in (\ref{2dVaction}), 
indeed we have fixed the ``gauge'' by fixing the
background metric (\ref{bame}).
Now that action, with the two constraints, is an effective theory 
of two fields propagating in a two-dimensional, flat spacetime.
After imposing (\ref{bame}), the stress tensor of (\ref{2dVaction}) reads
\beq
2\, T_{ab}=-\pa_a\eta\pa_b\rho+\frac{1}{2}\pa_c\eta\pa^c\rho\gamma_{ab}-\frac{V(\eta)}{2}e^{-2\rho}\gamma_{ab}-\frac{\pa_{a}\pa_b\eta}{2}+\frac{\gamma_{ab}\Box_\ga\eta}{2}.
\eeq
Notice that it is traceless on shell.
That suggests that, discarding the dilaton, and considering only the conformal factor of the
metric as an effective dynamical field, one gets a conformal invariant action. 
For our purposes it is convenient to rewrite
everything in lightcone coordinates $x^\pm=x^1\pm x^2$. The equations of motion
take the following form:
\beq\label{eqmoto}
\pa_+\pa_-\rho-\frac{\pa_\eta V(\eta)}{4} e^{-2\rho}=0, \qquad
\pa_+\pa_-\eta+ \frac{V(\eta)}{2} e^{-2\rho}=0 
\eeq
with the two constraints $T_{\pm\pm}=T_{11}+T_{22}\pm 2T_{12}=0$:
\beq\label{const}
\pa_+\pa_+\eta+2\pa_+\rho\pa_+\eta=0 \qquad        \pa_-\pa_-\eta+2\pa_-\rho\pa_-\eta=0.
\eeq
We have a system of two coupled field equations 
and all the geometric data of the $(r,t)$-plane are encoded in the
Liouville field $\rho$,  ($\gamma=(-1,1)$ for the coordinate $x^{1,2}$).
Notice that, the latter equation of motion in (\ref{eqmoto}) plus the constraints (\ref{const}) implies the
former in (\ref{eqmoto}).
It is interesting to notice that the constraints do not depend on the dilatonic potential,  
and so they are the same as for the theory in every dimension.

\s{Near-Horizon Approximation and Black Holes}

Up to now we have made the ansatz of spherical symmetry only. 
In this section we consider a spacetime with a black hole, in particular
we shall analyze the consequences of its presence on the equations of motion.
As pointed out above, imposing the constraints (\ref{const}), the equations of motion (\ref{eqmoto}) are not
independent.
That fact suggests that the solutions of the equations of motion which involve two fields, actually may 
be written in terms of a single field.
Notice that even in the case of CGHS model \cite{cghs} there is only one free field responsible for the whole 
dynamics of the black hole \cite{ns98back}.  
The standard procedure requires to fix the constraints and afterwards to integrate the equations of motion, but that
is not straightforward in the case of a general potential $V(\eta)$. 
On the other hand, it seems more natural to study a solution of the equation of motion near the horizon
 of a black hole, 
then to consider fluctuations of one of the two fields of the 
effective action (\ref{Ieff}) around it\footnote{For instance Solodukhin \cite{solo99} considers only
fluctuation of the dilaton $\eta$.}.  
Moreover those fluctuations shall not change the thermodynamical behavior of the black hole in question.

Therefore we discard for the moment the action, and we study the solutions of equations of
motion of a two-dimensional black hole.
In particular we consider some fluctuations of the field $\rho$ around the black hole 
solution\footnote{Some other authors consider a scalar field propagating in a
black hole metric finding a near-horizon CFT \cite{padm02}.}.
Notice that, 
the dilaton takes the same value on the horizon of every non-rotating black hole with a particular temperature
and entropy\footnote{See for instance \cite{cghs}.}.
Whereas there are variations of the conformal factor which preserve 
thermodynamics describing the black hole. (For instance by multiplying the metric
by a constant, the temperature and the entropy of the black hole do not change).
Below we shall show those facts in details performing the near-horizon limit.

Consider a bifurcate horizon of a black hole solution of the equation of motion (\ref{eqmoto}).
As usual there is a frame of lightcone coordinates $x^\pm$ which covers the part of the spacetime outer of the
horizon. In this coordinate patch the future/past-horizon is located at
$x^-=+\infty$/$x^+=-\infty$, and the conformal factor tends to zero on the horizon.
Let us perform the near-horizon limit of that solution.
It follows, in a straightforward way, that the near-horizon limit ($x^\pm\to\mp\infty$) we are considering
is equivalent to take the limit $\rho \to \infty$ (See section 5 for further details in the case of a 
near-horizon 
approximation of a Schwarzschild black hole by means of Rindler one.). 

If we integrate the constraints (\ref{const}) we gain the following expression:
\beq
\pa_\pm\eta=\exp{\at-2\rho+C_\mp(x^\mp) \ct}.
\eeq
Above $C_\pm$ are some finite function of $x^\pm$ respectively.
Performing the near-horizon limit of the constraints ($x^\pm\to\mp\infty$ and $\rho\to\infty$), we find that
the dilaton is almost constant near the horizon. 
Therefore, under that limit, we may consider the dilaton $\eta$ fixed at its value on the
horizon $\eta_0$.  
In that case, one can check by inspection that, 
constraints (\ref{const}) and the latter equation of motion in (\ref{eqmoto}) are satisfied.
Only the following equation for the Liouville field $\rho$ survives.
\beq\label{eqmotoredm}
4\pa_+\pa_-\rho-\pa_\eta V(\eta_0) e^{-2\rho}=0.
\eeq
Hence, at least in a region near the horizon of a black hole described by the fields 
$\rho_B$ and $\eta_B$, there is 
a solution $\rho_L$ of the equation (\ref{eqmotoredm}) that behaves as $\rho_B$ near the horizon.
In other words, the dynamics of the system is
completely described by the conformal factor $\rho$ near the horizon. 

In the particular case of a four-dimensional Schwarzschild-like spacetime
($V(\eta)=1/\sqrt{\eta}$), the ``near-horizon'' equation of motion
(\ref{eqmotoredm}) becomes: 
\beq\label{eqmotoliou}
\pa_+\pa_-\rho+\frac{1}{8 \eta_0^{3/2}} e^{-2\rho}=0.
\eeq
This is the equation of motion of a classical Liouville theory.
As is well known, the Liouville theory has a computable classical central charge.
In this approximation the equation of motion (\ref{eqmotoliou}) descends from an action of
the form
\beq\label{actionliou}
I=C\int_\Si dx\sqrt{\hat{g}} \at \frac{1}{2}\pa_\mu \rho \pa^\mu \rho +\frac{\mu}{\beta ^2}
e^{-\beta \rho} - \frac{2}{\beta}\rho \ca{R}[\hat{g}]\ct, 
\eeq
with $\beta=2$ and $\mu=-2V'(\eta_0)$ which, in the case of $V(\eta)=1/\sqrt{\eta}$, is
$\mu=1/\eta_0^{3/2}$.
We stress that we have not derived this action (\ref{actionliou}) from the gravitational one (\ref{Ieff}), 
but we have written an action that leads to the expected equation of motion (\ref{eqmotoliou}).
Jackiw in his well known paper \cite{jac84} follows the same procedure.
Moreover our theory has to be thought as an effective theory of the black hole. Perhaps, a more satisfactory
description of that assumption has to be sought in a possible quantum theory.

The factor $C$ of that effective theory differs from that presented above.
In the next we have to fix it in order to have the energy equal to the mass of the 
black hole\footnote{Another procedure, concerning micro-canonical action of general relativity, can be
found in \cite{bryo93}.}. To check this effective model of the bifurcate black hole, we 
shall compute the entropy and compare it with the thermodynamical one, finding a total agreement. 
Notice that we are studying a theory on a fixed background metric $\hat{g}$. 
Since we have extracted the conformal factor $\rho$,  we choose a flat background metric
$\hat{g}=\gamma=(-1,1)$.  
Therefore the term containing $\ca{R}[\hat{g}]$ vanishes.
It is well known that such a theory possesses a classical central charge, and to find  it we have to know the stress-energy tensor $T$.
In lightcone coordinates it has the following non vanishing components
\beq\label{stressl}
T_{\pm\pm}=C\at \pa_\pm\rho\pa_\pm\rho + \frac{2}{\beta}\pa_\pm\pa_\pm\rho \ct,
\eeq
notice that the last term arises from the variation of the action with respect to the metric 
before fixing the gauge, it is called improved stress-energy tensor. 
As is well known, the total classical central charge \cite{jac84} of this theory is
\beq
c=12C\frac{4}{\beta^2}.
\eeq
Notice that the value of the central charge computed above depends on the normalization constant $C$
of the stress tensor.
Moreover, both the central charge and the stress tensor do not depend on either $\mu$ and the particular value
of $\eta_0$. 
In this section we have derived the central charge of the gravitational action
performing a certain near-horizon approximation and considering the conformal
factor as the only dynamical field.
Notice that this central charge is peculiar to the black hole. 
We are considering all the fluctuations of
the conformal factor as degrees of freedom of the classical theory.
At this level some physical question arises, for example  it would be interesting to understand how to
interpret the fluctuations from a geometrical point of view.

\s{The Virasoro algebra.} 

In this section we try to construct the full Virasoro algebra, and then, using the Cardy formula \cite{cardy},
 we compute the entropy of the effective theory describing the black hole.
We have already noticed that variations of the conformal factor correspond to conformal transformations of the
action of the black hole.
As usual, the generators $L_n$ of the conformal transformations arise from the stress tensor
$T_{\pm\pm}$, and are computed smearing the stress tensor by some function.
\beq\label{charges}
L_n^\pm(x^\pm) = \int_S dx^\pm \xi^\pm T_{\pm\pm},
\eeq
where $S$ is the set that contains the $x^\pm$.
The $L_n$ are called the charges (Noether current) of the symmetry
in question. 
A general charge is composed of two parts corresponding to a right and a left moving plane waves
$\xi^+(x^+)+\xi^-(x^-)$ on the fixed space with metric 
${\ga}$. 
The solution of the Liouville equation (\ref{actionliou}) is $\rho=\rho^+(x^+)+\rho^-(x^-)$, the
$T_{++}$ and $T_{--}$ depend respectively only on $x^+$ and $x^-$.  
The usual Poisson bracket of the charges (\ref{charges}) satisfy the following relations\footnote{For
further details see \cite{jac84} for instance.} 
\beq\label{bcpoisson}
\ag L_n^{\pm} ,L_m^{\pm} \cg_{\ca{PB}}  =L^{\pm}[\xi_n, \xi_m]+ \frac{c^{\pm}}{12}\Delta (\xi_n ,\xi_m), \qquad 
\Delta = \int dx^\pm \aq \xi^\pm _n \pa_\pm ^3 \xi^\pm _m - \xi^\pm _m \pa _\pm ^3 \xi^\pm _n\cq,
\eeq
where the bracket $[,]$ is the usual Lie bracket of fields $\xi_n=\xi^+_n+\xi^-_n$.
Notice that  we have two classes of Virasoro generators $L_n^\pm$ they form two independent algebras. 
The total central charge is given by the sum of the central charges of the two algebras
\beq
c =c^+ + c^-.
\eeq
In order to find a Virasoro algebra we must specify the form of 
$\xi^\pm$. We set  
\beq
\xi^\pm _n =\frac{\ell}{2\pi}\exp \at-\frac{i\,n\,2\pi x^\pm}{\ell}\ct
\eeq
where the factor $\ell/2\pi$ is necessary in order that the vector fields $\xi^\pm$ closes with respect to the
Lie brackets and $\ell$ is a cutoff parameter. 
With that prescription the generators become  
\beq\label{caricheadim}
L_n^\pm=-\frac{\ell}{2\pi} \int_{- \frac{\ell}{2}}^{\frac{\ell}{2}} dx^\pm  \, \exp \at\frac{-i\,n\, 2\pi
x^\pm }{\ell}\ct T_{\pm\pm}.
\eeq
The $L_n^\pm$ of that form satisfy (\ref{bcpoisson}). By shifting $L_0$, it is possible to 
put it in the well known Virasoro form:
\beq\label{virasoro}
\ag L_n^\pm,L_m^\pm  \cg_{\ca{PB}}=i(n-m)L_{n+m}^\pm+i\frac{{c}^\pm}{12}(n^3-n)\de_{n+m}.
\eeq
This is true if the constant $C$ in $T_{\pm\pm}$ equals one.
In our case $C$ is not one therefore we shall rescale the Poisson brackets, as explained later, in order to close the algebra.
Before ending this section we want to spend some words on the meaning of the cutoff parameter $\ell$.
The reason for its presence is that we want to find a countable set of $L_n^\pm$ and this can only be  done by
smearing the stress energy tensor with the Fourier modes ($\xi_n$) on a compact space, and then considering the limit as the 
cutoff goes to infinity without changing the algebra.

\s{Counting the Microstates}

Let us consider an explicit solution of the equations of motion, and compute the associated  $T_{\pm\pm}$.
We shall give a possible description of the microstates responsible for the thermodynamics behavior of the black
hole. Therefore we consider a spacetime with a bifurcate Killing horizon. 
In general, the metric of a bifurcate Killing horizon generated by a black hole has the form
\beq\label{2dmetric}
ds^2 =- A({r})dt^2+ \frac{1}{A({r})}d{r}^2+r^2d\Om^2,
\eeq
where $d\Om^2$ is the metric of the unit sphere.
If we consider the near-horizon approximation ${r}\sim r_0$, the function $A(r)=A'(r_0)(r-r_0)$ and
$A'(r_0)=2\ka$, the metric of the $(r,t)$-plane becomes that of a Rindler space time with $r=r_0+\ka y^2/2$:
\beq\label{rindlermetric}
ds^2 =-\ka^2y^2dt^2 +dy^2.
\eeq
It can be brought in conformal form 
\beq\label{2dmetriclight}
ds^2 = -dx^+dx^- \exp (x^+-x^-) 
\eeq
where the new coordinates are $x^-=\ka t -\log y$ and $x^+=\ka t +\log y$. This metric, as expected, is 
conformally flat and the conformal factor is 
\beq
-2\rho = x^+-x^-=2\log y.
\eeq
The past horizon is located at $x^+=-\infty$ whereas the future horizon is located at $x^-=+\infty$.
In these coordinates, the fields $\xi_n$ become
\beq
\xi_n^\pm\pa_\pm=\xi_n^\pm\at \ka \pa_t\pm y\pa_y \ct.
\eeq
It would by nice if these fields, thought as generators of symmetry transformations, preserve 
the temperature and the entropy of the black hole in question.
To check this, we recall the expression of the temperature $T_h$ of the black hole,
\beq
T_h=\frac{1}{2\pi}\frac{\pa_y N}{\si}(r_0)
\eeq 
where $N$ is called the lapse function of the metric and $\si$ is the factor of
the spatial part $ds^2=-N^2dt+\si^2dy^2$, for the metric (\ref{rindlermetric})
$N=\ka y$ and $\si=1$. 

${\pa_y N}/{\si}=\ka$ is the surface gravity of the black hole which
is proportional to the temperature. We want to check if the variation of the temperature, namely the surface
gravity is zero under a transformation generated by $\xi^\pm_n$.
Moreover, we want to check if the horizon of the black hole is preserved, we recall that the horizon is located
where the lapse function vanishes. If  its variation vanishes, $\de N=0$, the variation of the
spatial area of the horizon, which is proportional to the entropy of the black hole, vanishes too.

We have the following Lie derivatives
\beq
\ca{L}_{\xi_n^\pm} N(r_0)=0, \qquad \ca{L}_{\xi_n^\pm} \frac{\pa_y N}{\si}(r_0)=
\ka \xi_n^\pm \aq \at in\frac{2\pi}{\ell}+1\ct^2-\at in\frac{2\pi}{\ell}+1\ct \cq.
\eeq
The variation of the lapse is always zero, and the variation of the surface gravity is zero in the limit of
large $\ell$.
Our fluctuations of the conformal factor of the metric, described by the field $\rho$, in the limit of large
$\ell$ preserve the temperature and the entropy of the black hole, this means that we are computing a class of
metrics compatible with the same black hole (fully described by its temperature and entropy).
However the conformal transformations in question is physically meaningful only when the cutoff $\ell$ tends
to infinity. (The walls of the box tend to infinity.)

Notice that we have a classical Liouville field theory on a fixed flat background. We shall fix the energy of
this effective model equal to the energy of the black hole, namely its ADM mass. 
In that way it is possible to fix
the normalization of the action. 
The ADM mass counts the energy of a spacetime endowed with a black hole as seen by an observer located at
infinity. It sounds natural that an effective theory describing that black hole should have the same energy.
We stress that the stress tensor $T_{\pm\pm}$ we are considering is also an
effective stress tensor and not the gravitational one.
On the other hand the Rindler metric is a near-horizon approximation of any black hole with a bifurcate
Killing horizon.
As just pointed out, to fix the energy of the theory, we fix the constant $C$ of the stress tensor
(\ref{stressl}) 
as follows
\beq
C=\frac{\ka A_h}{2\pi\ell},
\eeq 
where $A_h$ is the spatial area of the horizon and $\ka$ is the surface gravity.
The energy of the system, computed by that factor $C$, equals the mass of the
black hole $M_B$, namely since $T_{11}=T_{22}$ 
\beq
\int_{-\ell/\sqrt{2}}^{\ell/\sqrt{2}} T_{11} dx_2=
\frac{1}{4}\int_{-\ell/2}^{\ell/2} T_{++}+T_{--}dx^++\frac{1}{4}\int_{-\ell/2}^{\ell/2} T_{++}+T_{--}dx^-=\frac{\ka
A_h}{8\pi}=M_B.
\eeq
As the result does not depend on the value of $\ell$, it holds when the cutoff parameter tends to
infinity.
Moreover if we rescale the coordinates by a factor $1/a$, $\tilde{x}=x/a$, the energy of the theory does not
change.  
In particular, $\tilde{\pa}_\pm{\tilde{\rho}(\tilde{x})}=\pa_\pm\rho(x)$ and 
$\tilde{\pa}_\pm\tilde{\pa}_\pm{\tilde{\rho}(\tilde{x})}=a\pa_\pm\pa_\pm\rho(x)$. 
Moreover the vectors
$(\tilde{\xi}^\pm(\tilde{x})/a)\, \tilde{\pa}_\pm=\xi^\pm(x)\pa_\pm$.
Discarding the ``tildes'', the new charges $L_n^\pm$ become 
\beq
L^\pm_n=\frac{A_h\ka }{2\pi\ell}\int_{-\ell/2a}^{\ell/2a} dx^\pm \, a \at \frac{\ell}{2a\pi}\ct \exp{\at -i \frac{2a\pi}{\ell} n  
{x}^\pm \ct}
\at\pa_\pm \rho \pa_\pm \rho+\frac{1}{a}\pa_\pm\pa_\pm\rho\ct .
\eeq
Since $C$ is not equal to one, we have to rescale the Poisson brackets in order to have a Virasoro algebra.
Notice that the charges suited above satisfy the following Virasoro algebra
\beq
\ag L_n^\pm,L_m^\pm  \cg=i(n-m)L_{n+m}^\pm+i\frac{{c^\pm}}{12}n^3\de_{n+m},
\eeq
where the brackets computed above are the usual Poisson bracket $\{\cdot,\cdot\}_{\ca{PB}}$ of
(\ref{virasoro})
 rescaled by a factor $-1/C$, namely $\{\cdot,\cdot\}=-(2\pi \ell)/(A_h\ka) \{\cdot,\cdot\}_{\ca{PB}}$.
In order to explicitly write the influence of $\ka$ of the (\ref{rindlermetric}) in the charges 
we choose  $a = \ka$, getting the background metric $ds^2=-\ka^2dx^+dx^-$.
Notice that the time $(x^++x^-)/2$ is equal to the Rindler time $t$ of (\ref{rindlermetric})
and the Euclidean Rindler time $\tau=-i(x^++x^-)/2$ is periodic with period $2\pi/\ka$.
In the new coordinates the generator $L_0^\pm$ becomes
\beq
L^+_0 =\frac{A_h \ell}{16\pi^2}
\eeq
and the corresponding central charge reads
\beq
{c}^+ = \frac{3A_h}{2\ell}.
\eeq
We are eventually able to compute the entropy of this theory using the logarithm of the Cardy formula
\beq
S=2\pi \sqrt{\frac{{c^+}L_0^+}{6}} = \frac{A_h}{4}
\eeq 
That is nothing but the Bekenstein-Hawking entropy associated with the bifurcate Killing horizon (\ref{2dmetric})
for every value of the cutoff parameter $\ell$.

The charges $L_n^{\pm}$ diverge as $\ell$, and the central charge tends to zero as
$1/\ell$ for large value of $\ell$, therefore their product do not depend on $\ell$. So we are able to
compute the black hole entropy by using the Cardy formula.
We recall that a zero central charge was also found in \cite{koga, hotta00}.
We have used only one copy of the Virasoro algebra, that corresponding to the future horizon, because the 
event horizon of the physical black hole is given by the future horizon.

\s{Discussion and open problems.}

We have studied the conformal factor of a dimensional reduced theory in the presence of a black hole with a
bifurcate Killing horizon.
In this case we have shown that near the horizon, it is described by a Liouville theory, which does not modify the
thermodynamical properties of the black hole. This has a classical central charge and, in fact, we have shown
that it can be used to compute the entropy of the black hole in question. 
If the black hole is extremal it does not posses a bifurcate Killing horizon and in this case the geometry of
the $(r,t)$-plane is not that of a Rindler space but it is an $AdS_2$ space.
In this case our approximation does not hold because 
the exponential term in the equation (\ref{eqmotoliou}) is
positive, and so it is not a Liouville equation.
On the other hand, the temperature of an extremal black hole is zero.

We stress the fact that we have  made no assumptions on 
a specific quantum gravity model but used only the classical near-horizon structure of the black hole.
That approach differs from the already studied models because the central role is played by the conformal
field $\rho$ and not by the dilaton $\eta$. 
The effective conformal theory found above describes the micro-canonical theory responsible for the entropy of
the black hole, indeed we have fixed the value of $\int T_{tt}dr$ as its mass $M_B$.
We have found the correct thermodynamical entropy even if the central charge is zero if computed on the global
space, and even if the fundamental mode $L_0^+$ diverges as the cutoff parameter $\ell$ tends to infinity.
\noindent
The charges, if computed on the whole space time, correspond to conformal transformations that do not change
the thermodynamical properties of the black hole. 
On the other hand we have made no assumptions on the boundary condition we have to impose to the conformal
factor. Perhaps some other correction should be searched in that direction.

\section*{Acknowledgments.}

The authors are grateful to  L.Vanzo, V.Moretti, S.Zerbini and M.M.Caldarelli for useful discussions on the subject.


\end{document}